# Clustering of anterior corneal surface shapes in normal adults

Hala Bouazizi[1,*], Isabelle Brunette[2,3], Jean Meunier [1]

1 Department of Computer Science and Operations Research, University of Montreal, Montreal, Quebec, Canada
2 Department of Ophthalmology, University of Montreal, Montreal, Quebec, Canada
3 Maisonneuve-Rosemont Hospital Research Center, Montreal, Quebec, Canada

*Correspondence: hala.bouazizi@umontreal.ca

## Abstract

In the present study, we investigated a large dataset of normal adult corneal topographies in an exploratory attempt to identify the natural groupings of their 3D shapes, with the practical clinical aim of categorizing the shape of a future generation of biosynthetic corneal implants. Because of the wide range of shapes among normal corneas, we needed to identify a limited number of normal corneal shape categories to guide the conception of different implants to match the patient's own corneal shape category. At this stage of our research, we focused on the anterior surface, which exhibits greater variation in a normal population and is responsible for most of the refractive power of the eye. The corneal surface 3D data used were the anterior elevation maps taken from corneal topographies. Clustering corneal topography elevation maps was performed to reveal the normal corneal shape categories. To facilitate groupings, corneal data had their dimensionality reduced by Zernike polynomial modeling. The resulting clusters were evaluated using different clustering scores ($SPR^2$, $R^2$), representations (coefficient polar charts, coefficient scatter plots, average elevation maps, average profile cuts) and statistical cluster comparisons (on different clinical measures). While several hard and soft linear and nonlinear clustering methods were tested in this way, k-means proved to be sufficient for this task. The best number of clusters was three and they were primarily differentiated according to corneal curvature (Zernike defocus coefficient). These clusters were related to established clinical parameters, confirming that the algorithm extracted relevant information based solely on the normal 3D shape of corneas.

## 1. Introduction

The cornea is the transparent structure forming the anterior wall of the eye. Corneal shape plays a critical role because it determines most of the eye's total refractive power and, consequently, visual acuity [1–3]. To measure the 3D shape of the corneal surface, we used corneal topography, the medical standard for corneal surface imaging. Traditionally, classification or grouping of corneas based on their 3D shape has been performed manually through expert selection of clinical data and averaging of the corresponding corneal topography maps [4]. Such supervised categorization methods have been used, for instance, to identify age-related differences in normal eyes [4], for the staging of the progression of Fuchs endothelial corneal dystrophy [5], for the staging of keratoconus [4], and to analyze the impact of corneal surgery on corneal shape [4,6]. These applications demonstrate the informative potential of group-based 3D corneal shape models.

One highly innovative clinical application motivating the present research is the optimization of the production of biosynthetic corneal substitutes (artificial corneas) [7–9]. Biosynthetic corneal substitutes are designed as an alternative to human donor corneal transplantation and promise advances that may be difficult to achieve with human donor corneas. They offer numerous advantages over human donor corneas. Because they are intended to replace the anterior surface of the cornea, their surface shape parameters need to be defined to restore a normal corneal shape and functional vision in patients whose diseased cornea requires replacement with an implant, thereby improving visual acuity and quality of life.

Because of the wide range of shapes among normal corneas, a limited number of normal corneal shape categories need to be identified to enable selection of an implant corresponding to the patient's own corneal shape category, analogous to adapting shoe sizes to different foot sizes. As a first step, we need to identify the optimal number of normal corneal shape groups to categorize the 3D shape of the implant that each patient should receive.

The approach most widely reported in the literature for obtaining corneal shape models is to group corneas according to various clinical data, including anatomical parameters (e.g., axial length, anterior chamber depth, and white-to-white distance), refractive parameters (e.g., sphere, cylinder and axis, and spherical equivalent), and demographic parameters (e.g., age, sex, and ethnicity) [4,10]. However, this supervised approach requires sufficient labeled clinical data to achieve good accuracy; otherwise, its applicability may be limited. Moreover, the choice of the most relevant parameters for grouping is not straightforward.

The second approach is to use an unsupervised methodology, or clustering. In unsupervised categorization, manual labeling of clinical data is no longer needed (except for validation purposes), as the categorization is based directly on raw corneal shape data (here extracted from topographic maps). Only a few studies have applied unsupervised classification to corneal data, primarily with the aim of automating disease identification [11–14].

Yousefi et al. [11] and Hallet et al. [12] applied unsupervised clustering to a large number of clinical and topographic parameters (e.g., thinnest-point thickness, keratometry readings, index of vertical asymmetry), rather than to raw topography maps (i.e., (x, y, z) coordinates). These parameters were first reduced through feature extraction to a few latent variables. A clustering method was then applied to group the corneas into a predetermined number of clusters to automate the diagnosis of disease stages (in this case, normal, forme fruste, mild, and advanced keratoconus).

Zéboulon et al. [13] and Bouazizi et al. [14] directly applied clustering to raw topographic data instead of a set of clinical parameters, as described above, with a prior step of dimensionality reduction. Study [14] was based exclusively on normal corneas, whereas in [11–13], clustering was based on unlabeled normal and abnormal corneas, including keratoconus, Fuchs dystrophy, and post-refractive surgery corneas, with the aim of automating diagnosis.

The data in [13] included the elevations of the anterior and posterior corneal surfaces, as well as the inter-surface volume (corneal thickness or pachymetry). Feature extraction was based on t-SNE and reduced the initial set of $10{,}000 \times 3$ data points to only three eigenvalues. Clustering was performed using the HDBSCAN algorithm and the overall success rate was 96.5%, with better results for normal than for abnormal corneas. In [14], dimensionality reduction was achieved through geometric modeling using Zernike polynomials rather than feature extraction, and unsupervised classification of the corneas was based on agglomerative clustering of the anterior surface. In contrast to the previous studies [11–13], the optimal number of clusters in [14] was not prespecified but was determined from their degree of compactness and separation using a dendrogram and clustering scores. The four best clusters identified were assessed using average elevation maps and profile cuts representing their average shapes. The clinical variable that best reproduced the transformations in average corneal shape from one cluster to another, as visualized using average maps and profile cuts, was corneal curvature.

Bouazizi et al. [14] is the only study that has focused exclusively on raw topographic elevation data from normal corneas to identify their natural groupings. The present study further expands on these preliminary results. Here, we focus on the anterior corneal surface, which is responsible for most of the refractive power of the cornea [1–3]. The originality of our approach is to identify normal corneal categories solely on the basis of their 3D shape, without a priori clinical information (except for their normality). The resulting clusters of normal corneal shapes could provide a limited number of shapes for the development of new implants that are better suited to individual patients than traditional corneal grafts from eye-bank donors, for which no shape information is available. Once the most effective clustering method had been identified in preliminary tests and the clusters had been generated, their correspondence with significant clinical parameters was evaluated.

## 2. Method

### 2.1. Data

*Collection and description*

The investigated dataset comprises a total of 8,609 topographies of the anterior corneal surface from normal corneas obtained from 4,941 consenting adult subjects (4,245 female and 4,364 male corneal topographies, comprising 4,438 right-eye (OD) and 4,171 left-eye (OS) topographies). Some subjects contributed topographies from both eyes, whereas others contributed data from only one eye. These topographies were generated using the Orbscan II topographer (Bausch & Lomb, Rochester, NY) and are part of the Database for the Anatomopathological, Functional and Surgical Characterization of the Cornea of the Quebec Vision Health Research Network [15].

The database was screened by experienced ophthalmologists and trained technicians to ensure that subjects had no history of ocular disease or ocular surgery and no recent contact lens wear. Each corneal topography was recorded as a 101 × 101 grid of elevation values (heights along the $Z$ axis), evenly spaced at 0.1 mm intervals along the $X$ (left–right) and $Y$ (up–down) axes. Each topography was accompanied by demographic information about the participant (sex and age), geometric information about corneal shape (white-to-white diameter [WTW], anterior chamber depth [ACD], and eye side), and refractive information (cylinder, sphere, axis, and spherical equivalent). Only one topography per eye was used. The experimental procedures involving human subjects described in this paper were approved by the Institutional Review Board. Descriptive statistics for selected key clinical parameters in this study are provided in Table 1.

| | **Age** (years) | **Axis** (°) | **Cylinder** (Diopters) | **Sphere** (Diopters) | **SE** (Diopters) | **WTW** (mm) | **BFS R** (mm) | **ACD** (mm) |
|---|---|---|---|---|---|---|---|---|
| **Mean** | 40.26 | 81.12 | -0.62 | -3.13 | -3.44 | 11.84 | 7.91 | 3.69 |
| **Std** | 11.45 | 56.46 | 0.92 | 2.37 | 2.43 | 0.39 | 0.25 | 0.38 |

SE: Spherical Equivalent

Table 1: Descriptive statistics of some key clinical parameters of the dataset.

*Preprocessing*

It is not uncommon for some topographies to include a few artifacts, especially in the far periphery (e.g., missing data, noise, occlusion by the eye lids, lashes or nose). To address this issue, the matrices of elevation points were reduced into 91 × 91 matrices to eliminate 0.5 mm of radius and clean up the noisiest part of the surface (the image fringe). The resulting dimensionality of the data (8281 elevations), still substantial, could cause overfitting issues and was further reduced by geometric modeling using 16 Zernike polynomials (Zernike modeling), a number that provides sufficient accuracy [16] [17] for producing mean surfaces while avoiding overfitting issues.

*Zernike polynomial modeling*

Zernike polynomials (ZP) are known to be effective for describing optical aberrations of normal corneas [16] and for modeling corneal surfaces [16] [17]. They have the properties of linearity, completeness, orthonormality, and radiality, with clinically interpretable coefficients, which is an asset for interpreting clusters. A Zernike model of the corneal surface can be conceived as a matrix of elevations modeled as a sum of least-squares-fitted Zernike polynomials over the unit disk [16]:

$$S(\rho_w, \theta_w) = z_w = \sum_{j=1}^{J} C_j P_j(\rho_w, \theta_w) + \epsilon_w \quad \text{for } w = 1, 2, \ldots, W \tag{1}$$

In Zernike modeling, the corneal surface $S$ consists of a set of $W$ elevations $S(\rho_w, \theta_w)$ located at polar coordinates $\rho_w$ and $\theta_w$ for $w = 1$ to $W$. Each elevation $S(\rho_w, \theta_w)$ (equivalently $z_w$) is modeled by summing the products of the $j^{th}$ Zernike polynomial $P_j(\rho_w, \theta_w)$ and the $j^{th}$ coefficient $C_j$ for $j = 1$ to $J$, assuming error $\epsilon_w$. The $J$ Zernike polynomial terms were ordered according to the Noll sequential indices $j$ (see Table 2). The classical names of these functions for various values of $j$ are 1: piston, 2: horizontal tilt, 3: vertical tilt, 4: defocus, 5: oblique astigmatism, 6: vertical astigmatism, 7: vertical coma, 8: horizontal coma, etc. The first term (piston), which represents the mean height of the surface (as opposed to its shape), was not used or counted (having a much larger magnitude than the shape coefficients, its use in clustering pretests yielded irrelevant results based mainly on the mean height of corneal surfaces). $J = 17$ (16 coefficients without piston) was chosen as being accurate enough for the present purpose without causing overfitting (modeling error of RMSE = 2.98 µm in this dataset, which is lower than the repeatability error of the topographer of RMSE = 6.18 µm for a 91 × 91 matrix) [17]. Normalization over the unit circle implies that the topography polar coordinates $(r, \theta)$ must be scaled to the normalized polar coordinates $(\rho, \theta)$ by setting $\rho = r/r_{max}$, where $r_{max}$ denotes the maximum radial extent of the corneal topography region of interest ($r_{max} = 4.5$ mm in our study for the 91 × 91 elevation maps).

The definition of the Zernike polynomials is:

$$P_j(\rho_w, \theta_w) = \begin{pmatrix} \sqrt{2(n+1)}\, R_n^m(\rho_w)\, cos(m\theta_w)\,, j\ even, m \neq 0 \\ \sqrt{2(n+1)} R_n^m(\rho_w)\ sin(m\theta_w)\,, j\ odd, m \neq 0 \\ \sqrt{n+1}\, R_n^0, m = 0 \end{pmatrix} \tag{2}$$

where $n$ is the radial degree, $m$ is the azimuthal frequency, $j$ is the Noll index and

$$R_n^m(\mathrm{r}) = \sum_{s=0}^{\frac{n-m}{2}} \frac{(-1)^s (n-s)!}{s!\left(\frac{n+m}{2}-s\right)!\left(\frac{n-m}{2}-s\right)!} r^{n-2s} \tag{3}$$

The association between the Noll indices $j$ and the Zernike indices $(m, n)$ is provided in Table 2 for convenience. Figure 1 gives a color representation of the first Zernike polynomials, up to order four, with associated indices. Other details on Zernike polynomials can be found elsewhere [16].

| j | 1 | 2 | 3 | 4 | 5 | 6 | 7 | 8 | 9 | 10 | 11 | 12 | 13 | 14 | 15 | 16 | 17 |
|---|---|---|---|---|---|---|---|---|---|---|---|---|---|---|---|---|---|
| n,m | 0,0 | 1,1 | 1,-1 | 2,0 | 2,-2 | 2,2 | 3,-1 | 3,1 | 3,-3 | 3,3 | 4,0 | 4,2 | 4,-2 | 4,4 | 4,-4 | 5,1 | 5,-1 |

Table 2. Association between Noll indices *j* and Zernike indices (*m,n*).

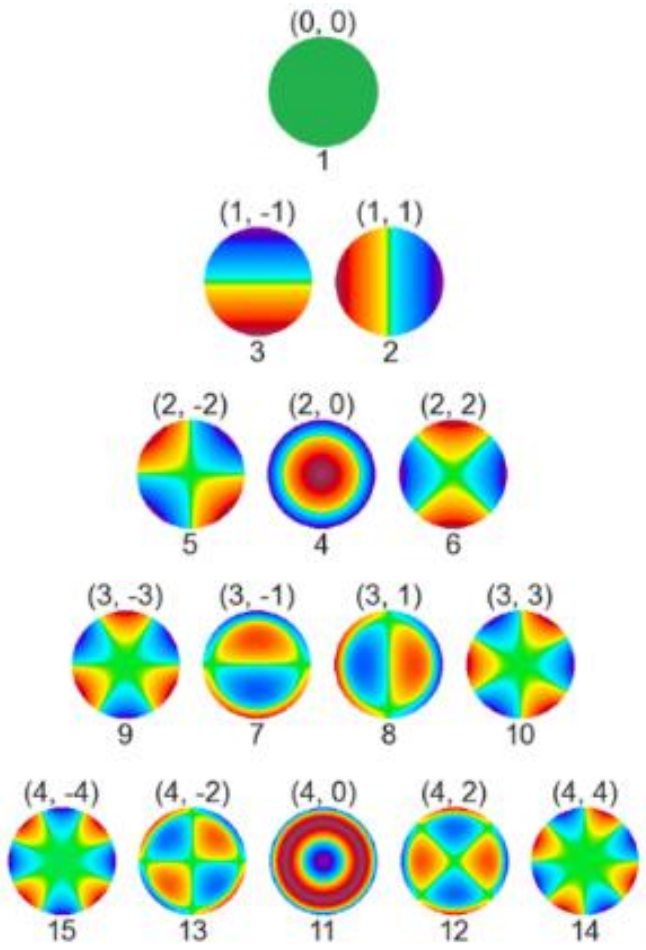

FIGURE 1: Pyramid of the Zernike basis functions up to order 4. For each Zernike function, *n* and *m* are above each elevation map and the Noll index *j* is given below. In this display positive values are cool colors and negative values are warm colors.

### 2.2. Clustering method

Several clustering methods were examined in pretests, including three linear methods (agglomerative clustering, k-means, and fuzzy c-means, made crisp by assigning each point to the cluster with the highest membership value) and three nonlinear methods (spectral clustering, the soft methods fuzzy c-means and Gaussian mixture model (GMM), also made crisp in the same way). A description of these standard algorithms can be found in [18]. Cluster evaluation was performed using the tools described in Section 2.3.

Data preprocessing was performed with or without geometric modeling (using raw elevation maps or Zernike polynomials) and with or without feature extraction (principal component analysis [PCA] for linear feature extraction or t-SNE for nonlinear feature extraction, as used in [13]). When necessary (e.g., for k-means, fuzzy c-means, and GMM), the algorithm was run 10 times with different centroid initializations to avoid suboptimal solutions, and the best result was selected. Other hyperparameters were optimized using various algorithms implemented in the Scikit-learn library [18] (e.g., the expectation–maximization [EM] algorithm for GMM), or several values were tested manually (e.g., the fuzziness parameter for c-means and the linkage method for agglomerative clustering).

For the soft clustering algorithms (GMM and fuzzy c-means), each data point (a corneal surface) was assigned to the cluster with the highest membership value. For nonlinear spectral clustering, we used a radial basis function (RBF) kernel for the affinity matrix.

Overall, geometric modeling with Zernike polynomials substantially improved clustering performance, as pretests without geometric modeling either failed because of computational costs or produced clusters that were unstable or difficult to interpret. Among the methods tested, k-means, fuzzy c-means, and agglomerative clustering performed consistently well, although inconsistent results began to appear with $k \geq 4$ (e.g., small, unbalanced, or heterogeneous clusters). However, fuzzy c-means performed better with low fuzziness, approaching the behavior of k-means (corresponding to fuzzy c-means with the lowest fuzziness parameter, = 1). Spectral clustering performance was suboptimal in our case. We suspect that the choice of affinity (RBF kernel) and the graph construction process had a greater impact than initialization. Alternative affinity matrices or kernel parameters could potentially improve clustering performance.

We also tested HDBSCAN (used in [13]), a density-based clustering algorithm that works best with data containing well-separated, high-density regions. In our case, because normal corneal shapes form continuous, smoothly varying distributions without clearly distinct high-density clusters, HDBSCAN did not perform well.

Regarding feature extraction methods, both a linear technique (PCA) and a nonlinear technique (t-SNE) were applied to the raw elevation-map data (91 × 91 matrices) for clustering normal corneas. However, neither method led to significant improvements. Linear clustering algorithms (agglomerative clustering, k-means, and c-means) continued to produce good results, whereas nonlinear clustering methods still yielded clusters lacking clear and coherent patterns under some conditions. These observations suggest that the geometric (Zernike) modeling step had already captured the essential structural information required for effective clustering.

In summary, these pretests showed that k-means was sufficient to achieve accurate clustering of normal corneas, and the results obtained with this method are presented next.

In k-means clustering, data are partitioned into *k* groups to minimize the within-cluster sum of squares (WCSS) (see below). Each sample is assigned to its nearest centroid, after which the centroids are recomputed. This process is repeated iteratively until convergence, typically when the assignments stabilize or the improvement in WCSS falls below a predefined tolerance. As mentioned above, k-means can be sensitive to initialization and consequently may produce a suboptimal solution. To alleviate this issue, we ran the k-means algorithm 10 times with different centroid initializations. The final result was the best of the 10 runs in terms of WCSS. In practice, we found no significant differences between the different runs.

### 2.3. Clustering evaluation

In this study, the success of the corneal clustering task was evaluated using different evaluation tools, including tools to identify the best number of clusters (clustering scores), and tools to validate the clusters based on Zernike coefficient display, surface visualizations and statistical cluster comparisons.

*Clustering scores*

The clear delimitation of clusters (compactness and separation) is largely dependent on *k*, the number of clusters produced. Selecting a too small number of clusters will result in underfitting while selecting a too high number of clusters will result in overfitting. Following Halkidi et al. [19], the $SPR^2$ and $R^2$ scores were used to determine the best *k* (curves shown in Figure 2).

1) $SPR^2$ (semi-partial $R^2$) measures directly the gain of compactness produced when a new cluster is added. The $SPR^2$ score can be computed as the difference between two successive within-cluster sum of squares (WCSS) scores as *k* increases divided by the total cluster sum of squares (TCSS) of the whole data set which measures the total distance between all samples and the global centroid. WCSS and TCSS are defined as follows:

$$WCSS_k = \sum_{i=1}^{k} \sum_{\boldsymbol{x} \in \boldsymbol{Cl}_i} (\|\boldsymbol{x} - \boldsymbol{u}_i\|^2) \tag{4}$$

$$TCSS = \sum_{x} (\|\boldsymbol{x} - \boldsymbol{M}\|^2) \tag{5}$$

$$SPR^2 = \frac{WCSS_{k-1} - WCSS_k}{TCSS} \tag{6}$$

where $Cl_i$ is the $i^{th}$ cluster, $\boldsymbol{u}_i$ is the centroid of cluster $Cl_i$, $\boldsymbol{M}$ is the overall mean of all samples and $\boldsymbol{x}$ is a sample from the dataset.

The $SPR^2$ curve is easier to interpret than the WCSS curve: the best $k$ is located at the elbow, where the curve becomes (roughly) flatter i.e. from a steep slope to a much flatter slope.

2) $R^2$ is a direct measure of cluster separation. It is estimated as the ratio of the between cluster sum of square (BCSS), which measures the average distance between the cluster centroids, and the total cluster sum of square (TCSS), which measures the total distance between all samples and the global centroid. The best $k$ is also located at the elbow (or knee) of the curve.

$$BCSS = TCSS - WCSS \quad (7)$$

$$R^2 = \frac{BCSS}{TCSS} \quad (8)$$

*Coefficient visualizations*

1) *Coefficient polar charts*. These charts were used to represent the mean coefficients of the corneal surfaces belonging to a cluster. In Figure 3, each line represents the means of the 16 coefficients of a cluster. The more distinct and separated the lines are for a given coefficient (from $C_2$ to $C_{17}$), the more this coefficient is discriminative and effective as a clustering criterion in the clustering process.

2) *Coefficient scatter plots*. Complementary to coefficient polar charts, coefficient scatter plots represent the distribution of the coefficients for each cluster for a given value of $k$. Figure 4 consists of a series of scatter plots that represent the distribution of the Zernike coefficients of each anterior corneal surface belonging to a cluster for every possible pair of coefficients ($C_2$ vs. $C_3$, $C_2$ vs. $C_4$, etc.). In the scatter plots, each cluster is assigned to a specific color. The more clearly the colors are separated across the scatter plots for a given Zernike coefficient, the more this coefficient is discriminant and effective as a clustering criterion.

*Surface visualizations*

1) *Average elevation maps*. These maps were used to represent the average surface of the corneas of each cluster (see Figure 5). An elevation map is a topographic representation of the elevations of a corneal surface. Assuming that the shape of the cornea is approximately spherical, the elevations are represented relative to the sphere that best fits the surface, the best-fit sphere (BFS). The elevations that are above the BFS are pictured with warm colors, and those that are below the BFS with cool colors. In an average elevation map, the surface represents the average of a group of corneal surfaces. The average surface was computed for each cluster by averaging the Zernike coefficients of the surfaces belonging to the cluster and reconstructing an average corneal surface on this basis (this is equivalent to computing the means of their elevations). A *cluster BFS* is the BFS of the average surface of a cluster. A *common BFS* is the average BFS of all clusters obtained by averaging their cluster BFSs. Because average elevation maps show the 3D shape of the cornea, they can serve to interpret the clustering results in terms of clinical parameters and Zernike coefficients. For each cluster, the mean elevation map was expected to present a coherent and interpretable pattern. Maps displaying erratic or heterogeneous color distributions were considered indicative of overfitting (an excessively large number of clusters, $k$) or an inappropriate clustering method in the pretests (section 2.2).

2) *Average profile cuts*. These representations are horizontal or vertical profile cuts of the average cluster surfaces (see Figure 6). They are computed as the horizontal or vertical meridian elevations of the average corneal surface of each cluster. For better visualization, the differences of the cluster surface cuts with the common BFS cut were amplified tenfold in our work.

*Statistical cluster comparisons*

Clusters were compared based on the clinical features known to be associated with the clustering criterion that best distinguished them. For example, if the clusters were found to be mainly distinguished by the Zernike coefficient defocus, they were compared with clinical features associated with it such as corneal curvature. Statistical comparisons between clusters were performed using analysis of variance (ANOVA).

## 3. Results

We present in detail clustering tests with k-means applied to OD corneal surfaces modeled with 16 Zernike coefficients. Comparisons with OS corneas are also provided for the surface visualization with average maps.

### 3.1. Clustering scores

Figure 2 presents the $SPR^2$ and $R^2$ curves for the k-means clustering of anterior OD corneal surfaces modeled with 16 shape Zernike coefficients. Using the elbow (or knee) technique, the best $k$ value for a clear delimitation should be $k = 3$. $k = 4$ was also investigated but was less stable across pre-test conditions (see Section 2.2 and Discussion) showing probably the beginning of some overfitting.

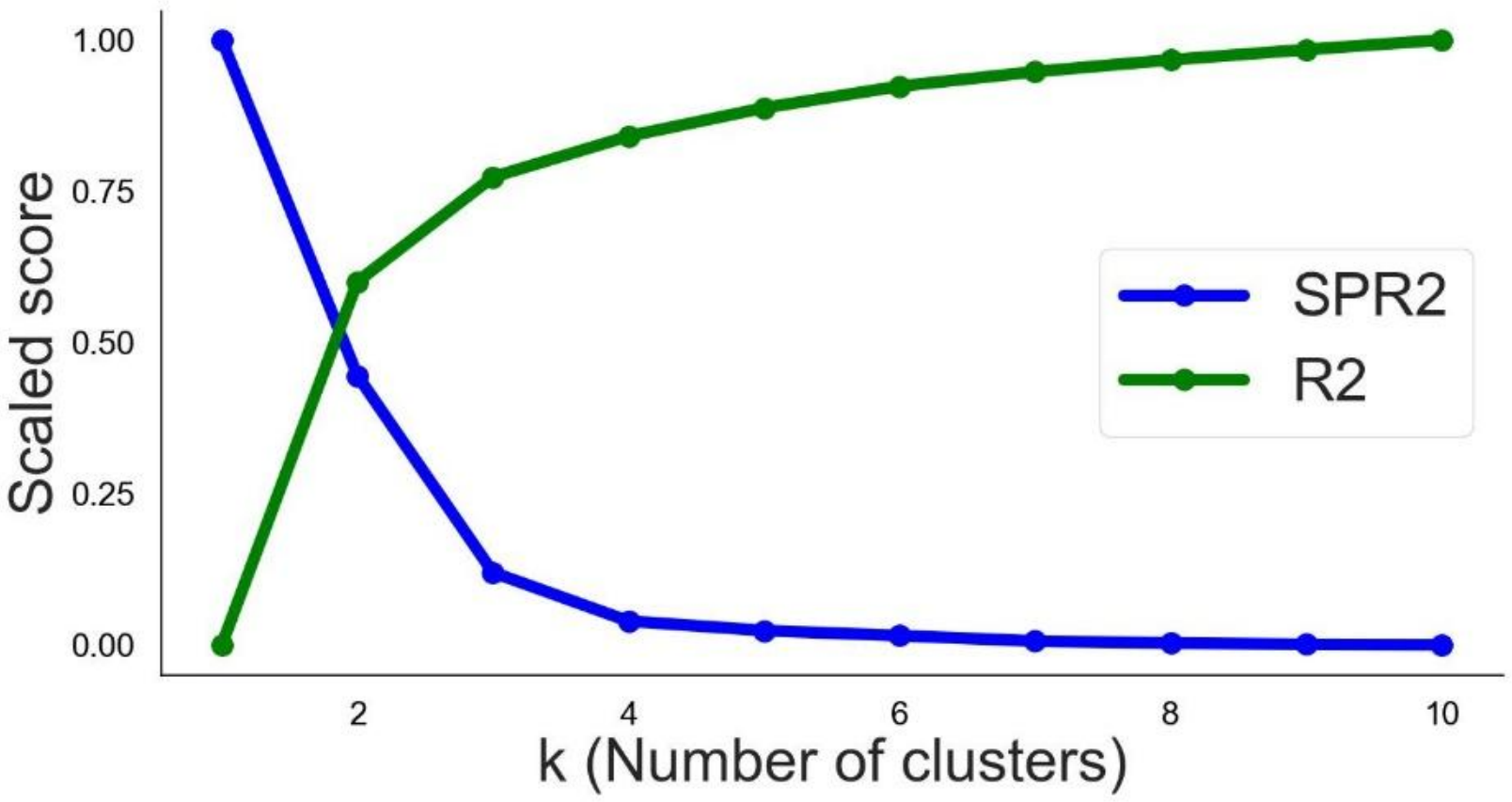

FIGURE 2: Clustering score curves for anterior OD corneal surfaces modeled with 16 shape Zernike coefficients.

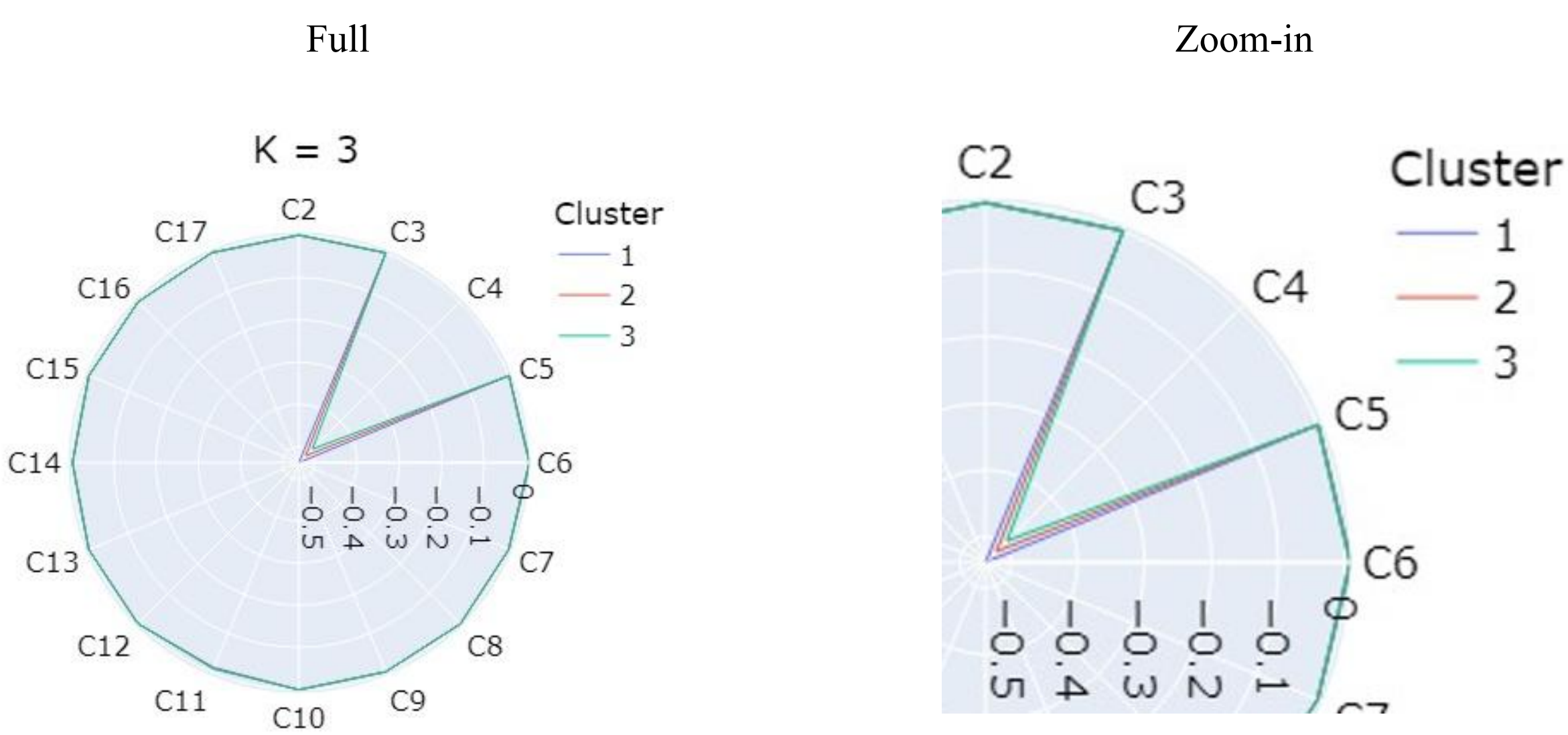

FIGURE 3: Coefficient polar chart based on 16 Zernike shape coefficients using k-means clustering with $k = 3$ for corneal anterior surfaces. (Left) Full view of the polar chart. (Right) Zoom-in view of the chart that focuses on the most discriminant Zernike coefficient.

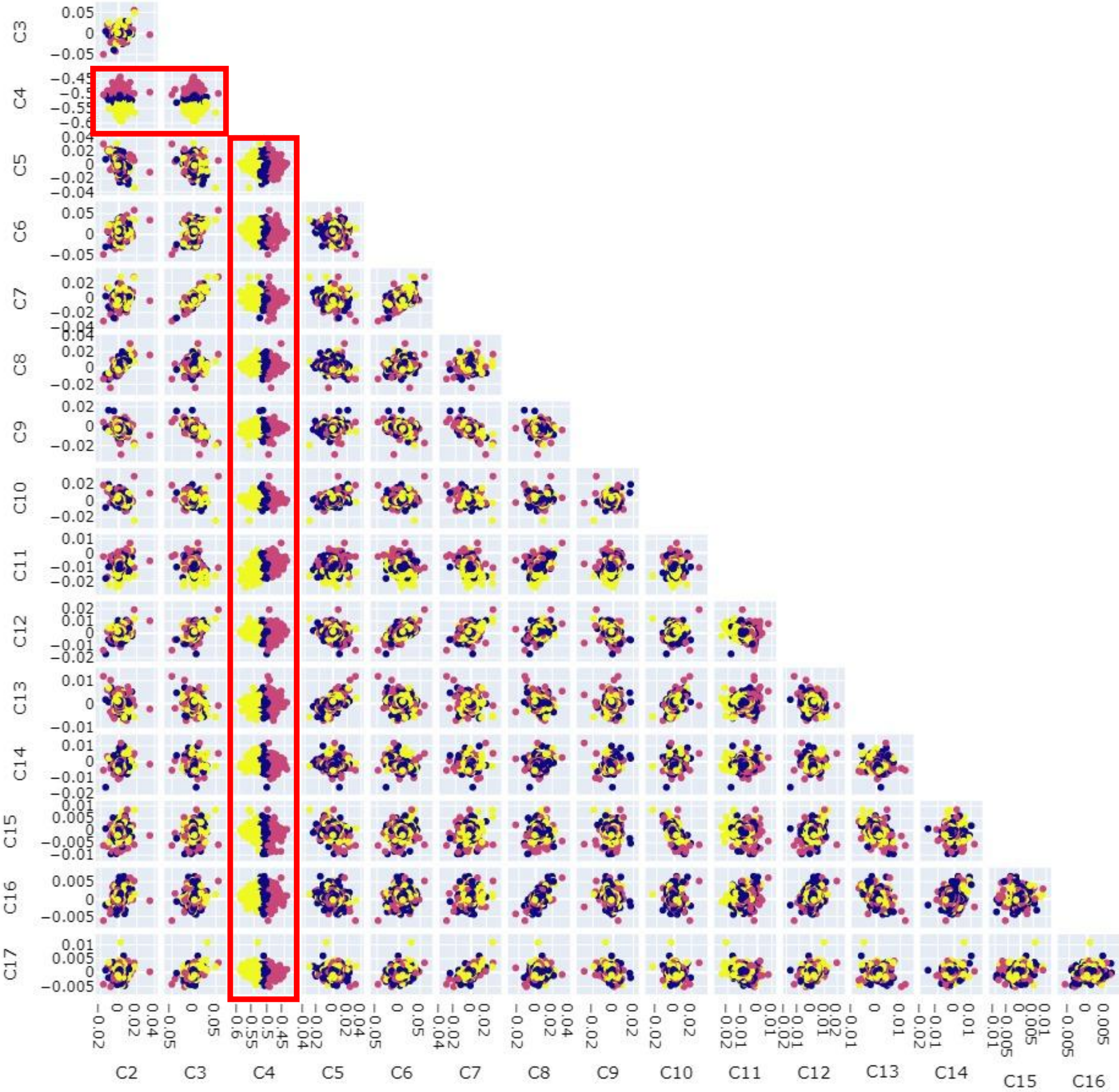

FIGURE 4: Series of pairwise coefficient scatter plots of the distribution of the 16 Zernike coefficients based on the OD anterior corneal surface clustering for $k = 3$.

### 3.2. Coefficient visualizations

*Coefficient polar charts.* Figure 3 presents the coefficient polar charts for the k-means clustering of anterior OD surfaces modeled with $k = 3$. A zoomed-in view of the chart focuses on the most discriminating Zernike coefficient, $C_4$. Indeed, $C_4$ is the only coefficient for which the $k$ cluster lines are clearly distinct. It is associated with the Zernike polynomial $Z_2^0$ shown in double-index notation and corresponds to defocus.

*Coefficient scatter plots.* Figure 4 presents a series of coefficient scatter plots showing the distribution of pairs of coefficients ($C_2$ vs. $C_3$, $C_2$ vs. $C_4$, etc.) for the clusters produced by k-means clustering with $k = 3$. As can be seen from the clearly delineated cluster colors, $C_4$ is by far the most discriminating Zernike coefficient for corneal anterior surfaces. It is also an order of magnitude larger than the other coefficients; the axis scales are not the same for all coefficients.

### 3.3. Surface visualizations

Average elevation maps show the 3D shape of the cornea and can be used to interpret the clustering results in terms of clinical parameters. When a clustering criterion aligns with a clinical parameter, we can expect to observe similarities between the cluster maps (the maps produced during the clustering process) and the atlas of that parameter (average elevation maps for a specified clinical parameter) [4]. Figure 5 shows a series of average elevation maps of the clusters produced by k-means clustering with $k = 3$ for OD surfaces.

In the first row, the original surface maps are referenced to the common BFS. As can be seen, they display a marked shift toward cooler colors at the center of the cornea as the BFS radius (R) increases from one cluster to another, indicating that the surfaces become progressively flatter from one cluster to another. We observe a very good similarity between these cluster maps and atlases based on BFS R and, to a lesser extent, those based on the white-to-white diameter (WTW), which is known to be highly correlated with BFS R [19-21].

In the second row, the original surface maps are referenced to their respective cluster-specific BFSs, showing similar anterior surface elevation patterns, with the typical positive central elevation (above the BFS, shown in yellow-orange) of a normal cornea. After normalization using cluster-specific BFSs, the effects of BFS size are largely eliminated, leaving only local features, such as a slightly more yellow (less orange) central region from one cluster to the next.

OD Anterior surfaces (in reference to the common BFS)

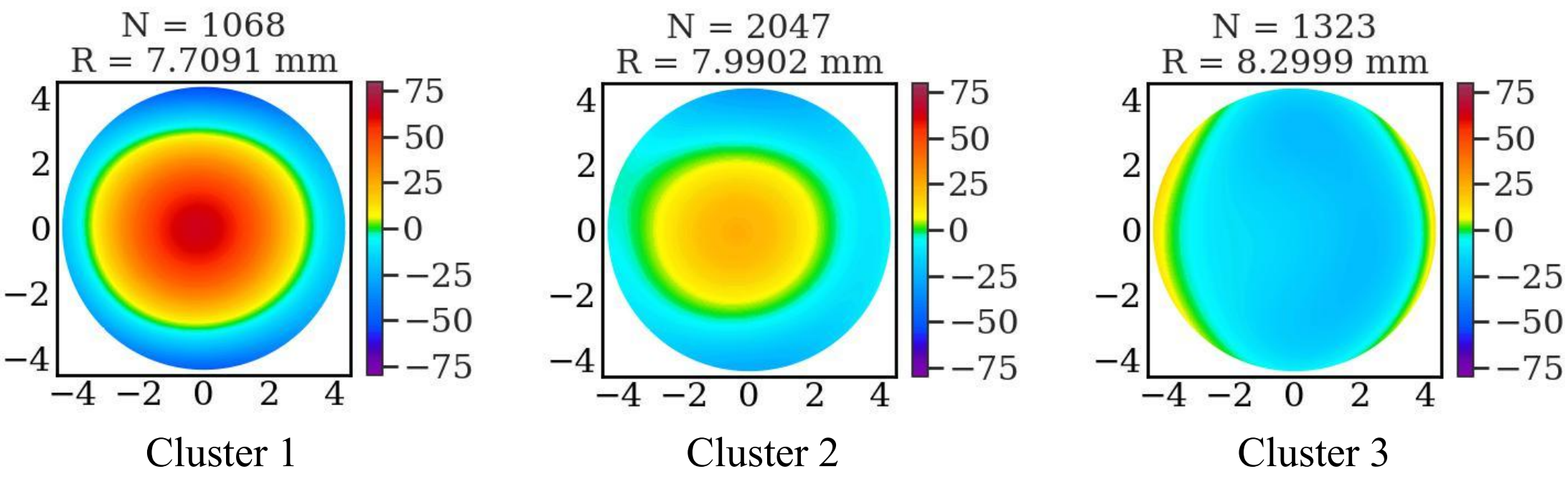

OD Anterior surfaces (in reference to their respective cluster BFSs)

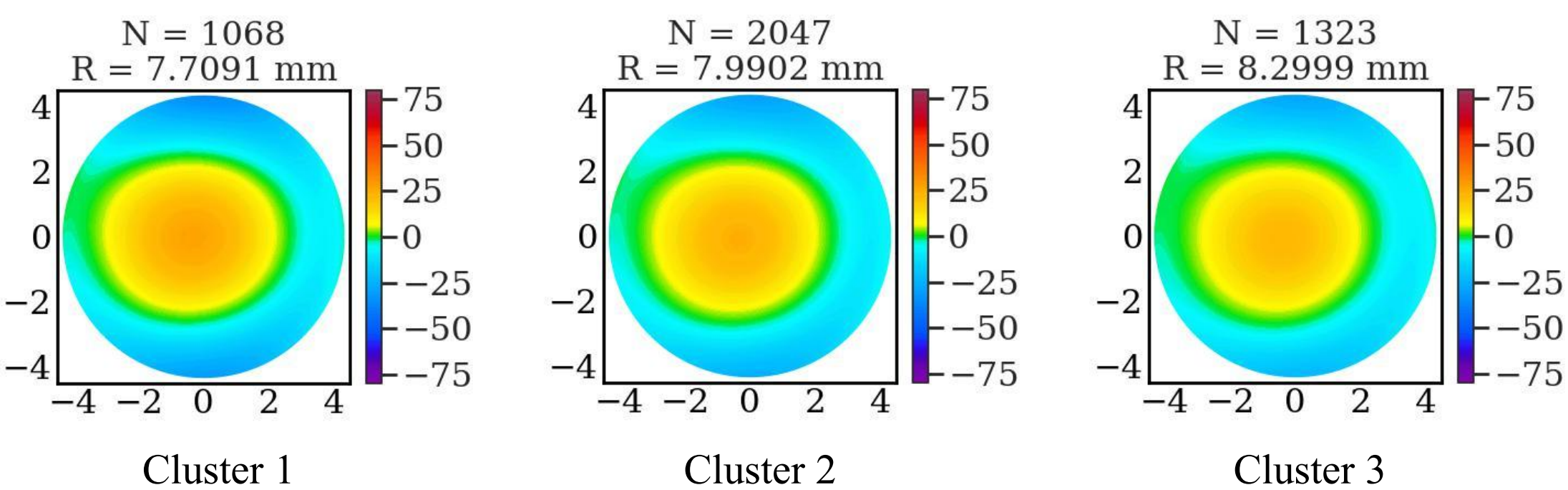

FIGURE 5: Cluster average elevation maps for k-means clustering for OD (right eye) surfaces modeled with 16 Zernike terms with $k = 3$. In the first row, the clusters are based on corneal anterior surfaces and the maps are referenced to the common BFS. In the second row, the clusters are referenced to their respective group reference BFSs. The $X$ and $Y$ axes are in mm, and the $Z$ axis (color scale) is in μm. R is the BFS radius of the average surface (elevation map) for each cluster, N is the number of corneas of each cluster.

*Average profile cuts.* The horizontal and vertical average profile cuts presented in Figure 6 referenced to the common BFS allow us to directly visualize the reduction in curvature from one cluster to another as the BFS R increases.

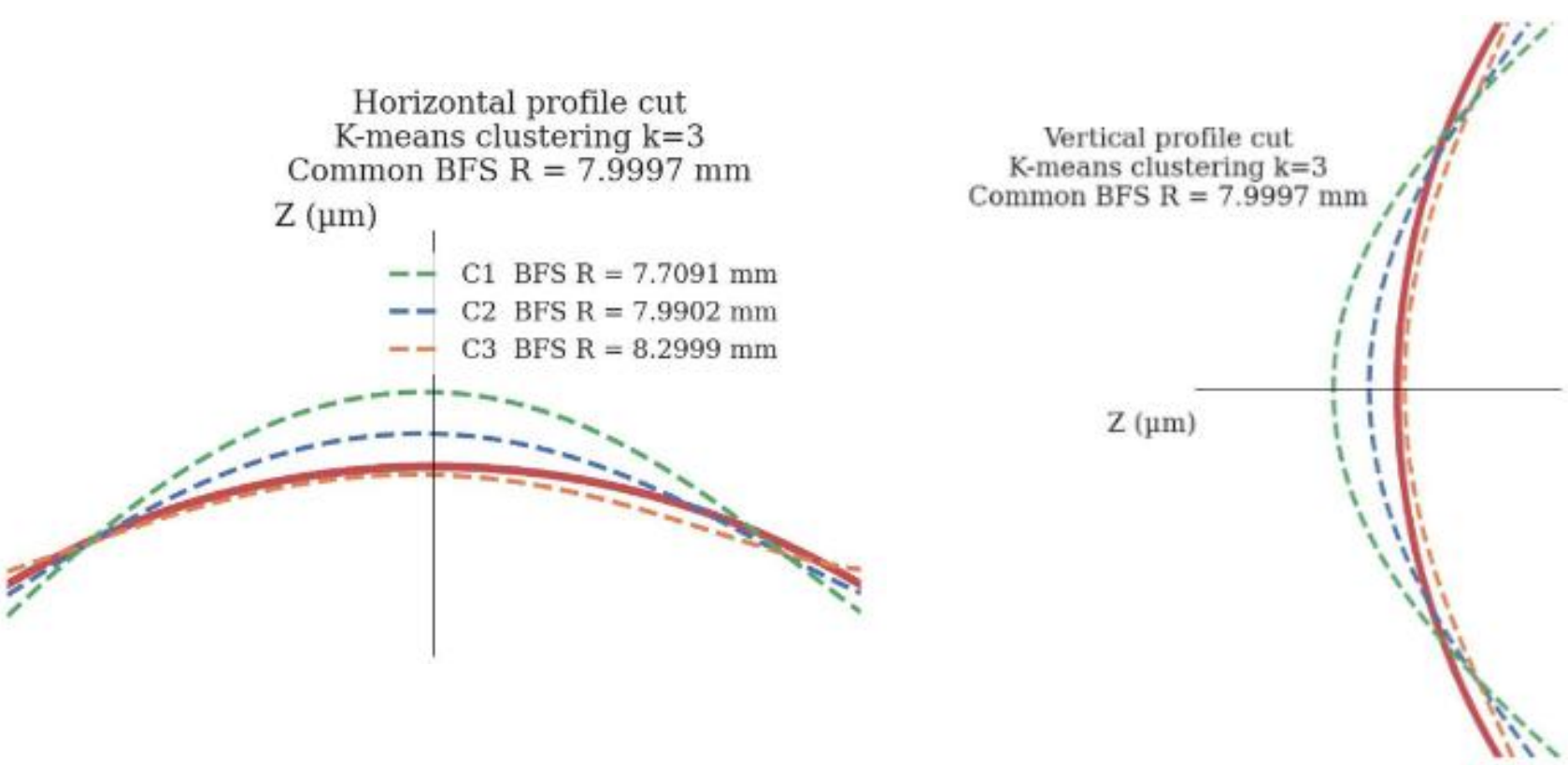

FIGURE 6: Average profile cuts along the vertical and horizontal meridians of OD corneal anterior surfaces modeled with 16 Zernike coefficients using *k*-means clustering for $k = 3$. The solid red curve is the common BFS. The differences of elevation between the mean cluster surfaces are amplified tenfold for better visualization.

### 3.4. Statistical cluster comparisons

As seen above, the unsupervised classification of anterior corneal surfaces produced by k-means clustering is mainly related to the Zernike coefficient of defocus ($Z_2^0$). The amplitude of the coefficient of defocus is related to the corneal curvature, known to be associated with some of the clinical features available in our database, including the BFS radius (R) [21], the white-to-white diameter (WTW) [20], the sex [10] [14] [21] [22] [23], etc. Therefore, we would expect the clusters to be associated with R, WTW, and sex. As shown in Table 3, this is indeed the case: the mean values of these variables increase or decrease progressively from one cluster to another as expected.[1] Each of these effects achieve statistical significance ($p < 0.0001$). The corresponding values of $C_4$ for each cluster are also provided for convenience.

Table 3: Effects of clustering methods on clinical variables

| Clinical Feature | Cluster 1 | Cluster 2 | Cluster 3 |
|---|---|---|---|
| R* (mm) | 7.60 ±0.12 | 7.89 ±0.09 | 8.21 ±0.15 |
| WTW* (mm) | 11.63 ±0.36 | 11.80 ±0.35 | 11.94 ±0.37 |
| Sex* | 0.66 ±0.47 | 0.50 ±0.50 | 0.36 ±0.48 |
| $C_4$* | -0.541 ±0.011 | -0.516 ±0.007 | -0.491 ±0.010 |

* $p < 0.0001$

[1] For sex, the score is the average of the values assigned to females (1) and males (0), so the higher the mean sex score for a cluster, the higher the proportion of females in that cluster.
The radius (R) values shown in Table 3 are the averages of the BFS radii computed for each individual surface within the corresponding cluster.

### *3.5. Replication of cluster average elevation maps with OS*

To evaluate the robustness and stability of the k-means clustering with $k = 3$, we replicated the cluster-average elevation maps shown in Figure 5 using a different dataset consisting of OS (left-eye) corneas from our dataset. If the clusters are reliable, the OS maps (Figure 7) should be similar to the OD maps (Figure 5), except for the expected enantiomorphism (mirror symmetry between the right and left corneas) [24]. As can be seen, this is indeed the case.

OS Anterior surfaces (in reference to the common BFS)

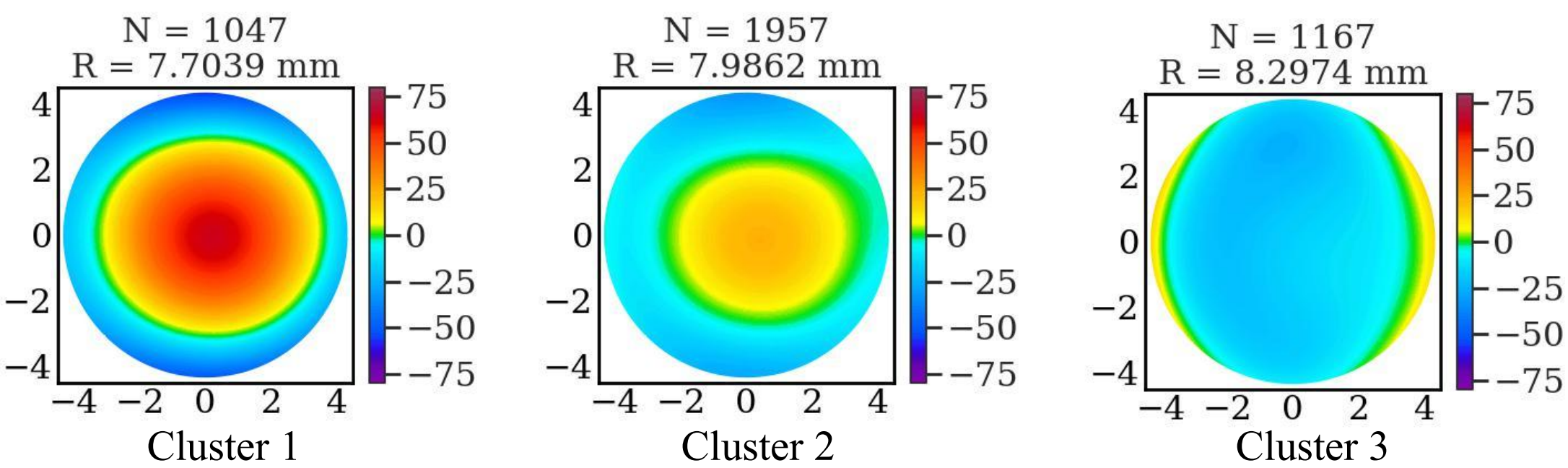

OS Anterior surfaces (in reference to their respective cluster BFSs)

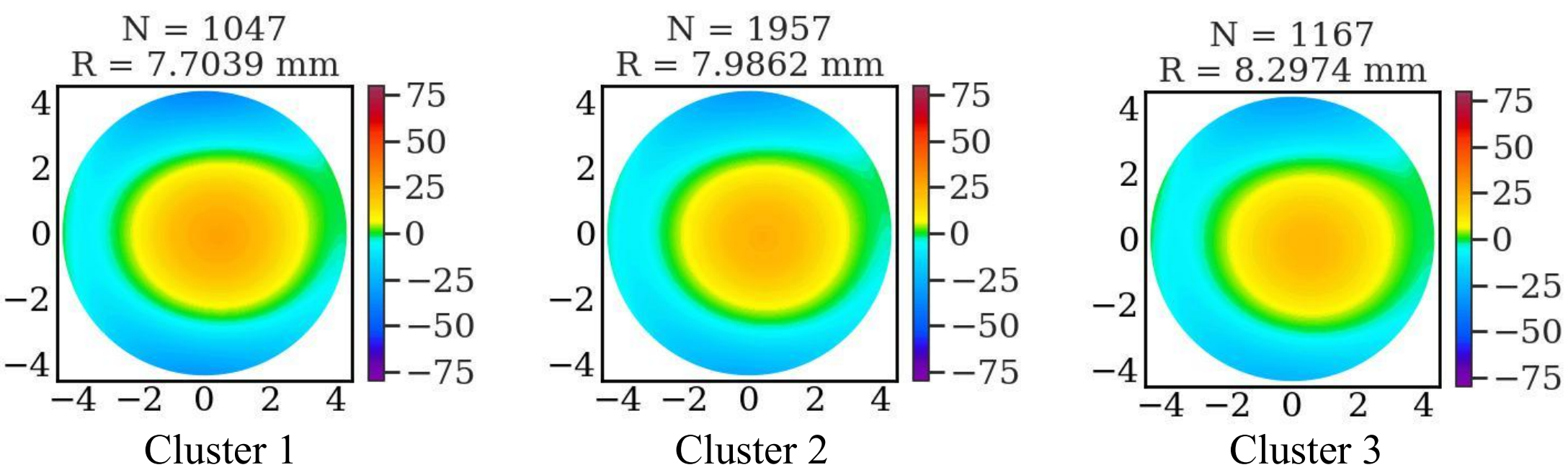

FIGURE 7. Cluster average elevation maps for k-means clustering for OS (left eye) surfaces modeled with 16 Zernike terms with k = 3. In the first row, the clusters are based on anteior surfaces and the maps are in reference to the common BFS. In the second row, the clusters are in reference to their respective group reference BFSs. The X and Y axes are in mm, and the Z axis (color scale) is in µm. R is the BFS radius of the average surface (elevation map) for each cluster, N is the number of corneas of each cluster.

## 4. Discussion

Contrary to prior studies that used unsupervised classification of corneal surfaces primarily to distinguish diseased corneas from normal ones [11–13], in this study, we examined the distribution of normal adult anterior corneal surfaces based solely on shape by testing clustering methods to identify natural groupings under optimal conditions. Corneal topography was used because it is the medical gold standard for 3D imaging of corneal surfaces [25–27].

Identification of the main categories of 3D corneal shapes in the normal population adds to our knowledge of corneal anatomy. However, the main clinical application of this study is the development of biosynthetic corneal substitutes [7-9, 28-31]. These substitutes offer numerous advantages over human donor corneas, including sterility, quality control, freedom from immune rejection, the possibility of controlling shape and refractive power, functionalization, prolonged storage, increased accessibility, and high-volume production [28]. Although the first biosynthetic corneal substitutes implanted in human patients showed highly promising results [28], postoperative topography showed steeper anterior curvature and greater surface irregularities than in healthy controls [29]. This finding underlines the importance of optimizing the surface curvature of these implants to minimize postoperative ametropia and astigmatism, an opportunity that is not available with traditional corneal transplantation using human donor tissue. With the new generation of implants requiring no sutures and not being subject to wound biomechanics, the shape of the implant's anterior surface becomes even more important. Construction of corneal implants according to 3D shape category could be achieved by casting, molding, or 3D bioprinting [30-31].

To restore the patient's cornea to a normal shape, we need to select an appropriate implant for the patient's ocular anatomy. We identified three categories of normal corneal shape based on topography, which could be used to design a limited number of implants adapted to individual patients, analogous to different shoe sizes. For example, a patient's BFS could be used to select the implant shape category with the closest corresponding BFS. If a patient's eye has a small BFS, the appropriate implant could be selected from the category of normal corneas with a similarly small BFS. The implant could then be shaped according to the average corneal shape of that category. BFS is generally robust enough to remain detectable in diseased corneas of patients who are candidates for biosynthetic corneal substitutes.

We focused on the anterior surface because it is primarily responsible for the refractive power of the cornea, has the largest change in refractive index, and consequently the greatest bending of light rays, occurs at the air–cornea interface (corneal refractive index ≈ 1.376; air = 1.0), whereas the change in refractive index between the cornea and aqueous humor (refractive index ≈ 1.338) is relatively small [3]. Focusing on the anterior surface also offers a significant advantage because this is often the only corneal layer that can be imaged in patients with diseased corneas awaiting transplantation, owing to the lack of transparency of the deeper layers. Nonetheless, for other applications, the same methodology could be used to categorize the posterior surface and pachymetry maps generated by the corneal topographer, thereby providing a more complete description of mean corneal shape.

While the dataset used in this study was comprehensive and consistent with the literature on anterior corneal shape across various demographic, refractive, and clinical anatomical parameters, it is important to acknowledge that the generalizability of the model to all populations may be limited. For instance, the dataset predominantly consisted of corneas from White participants, which may not fully represent the diversity of broader populations. It also contained more myopic corneas than would be expected in the general normal population (Table 1). Therefore, further validation using more diverse datasets is recommended to confirm the applicability of the model to different populations (e.g., Asian populations). In addition, the topographies were generated using the Orbscan II topographer (Bausch & Lomb, Rochester, NY), and other topographers could produce slightly different results.

Clustering was performed on normal adult anterior corneal surfaces using the simple but effective k-means algorithm with geometric modeling based on Zernike coefficients, without any additional feature extraction or transformation. K-means was efficient because the variance of the feature vectors in the normal population was primarily associated with the $C_4$ coefficient (defocus), suggesting that the clustering structure was low-dimensional. This finding explains why there was no apparent advantage in using sophisticated feature-extraction methods (PCA or t-SNE) or more complex clustering methods. Consequently, the elevation maps shown in Figure 5 represent the anterior surface shapes of potential small, medium, and large implants for OD (right-eye) corneal biosynthetic graft production, while Figure 7 shows the corresponding results for OS (left-eye) corneas.

Although defocus dominates, the other Zernike coefficients are not negligible and can provide meaningful information that is crucial for a comprehensive representation of the corneal surface. In fact, these other coefficients contribute 8% to the anterior surface modeling in Equation 1. Their importance can be observed in the second row of Figure 5 (or Figure 7), where the clusters reveal a slight temporal shift of the central bulge (yellow-orange region), with the temporal cornea known to be steeper than the nasal cornea [32]. This small off-center shift cannot be generated by defocus alone and likely reflects normal anatomical variations.

The optimal value of $k$ was 3 or 4 according to the $SPR^2$ and $R^2$ plots. Selecting too few clusters resulted in underfitting, whereas selecting too many led to overfitting. After several tests, inconsistent outcomes— such as small, unbalanced, or heterogeneous clusters—occasionally appeared with $k \geq 5$ when using linear clustering models (k-means, c-means, and agglomerative clustering). Nonlinear models tended to overfit earlier, with GMM showing signs of overfitting at $k = 3$ and spectral clustering at $k = 4$. To ensure greater stability and avoid overfitting, we therefore selected a linear model with $k = 3$.

Regarding cluster consistency, comparison of the OD (right) and OS (left) corneal clusters (Figure 5 vs. Figure 7) demonstrates cluster stability across two different datasets, as shown by their high similarity, apart from the expected enantiomorphism (mirror symmetry) between the right and left corneas [24]. Other tests using independent datasets also confirmed the stability of the three clusters for normal corneas. In our preliminary results reported in [14], agglomerative clustering with $k = 4$ was selected for normal adult anterior corneal surfaces; however, cluster stability was not evaluated, and the dataset was smaller and included both OD and OS corneas, with OS corneas flipped to resemble OD corneas. Agglomerative clustering was also significantly slower than k-means for a large dataset. Nevertheless, the cluster-average elevation maps were similar, with a four-level ($k = 4$) progression of average corneal shapes from one cluster to the next, rather than the three-level progression observed with $k = 3$.

Overfitting was primarily addressed by transforming the high-dimensional topography data (91 × 91 data points) into a low-dimensional representation using Zernike polynomial coefficients (16 coefficients), with no significant additional benefit observed from feature-extraction preprocessing. At this stage, for normal corneas, a higher number of coefficients or more advanced modeling techniques, such as spherical harmonics, were unnecessary because the modeling accuracy (RMSE = 2.98 μm) was better than the test–retest reliability of the topographer (RMSE = 6.18 μm) [17], and the clusters were well defined and interpretable.

Scatter plots and polar charts identified $C_4$ ($Z_2^0$) as the primary clustering criterion. Known as the defocus coefficient, $C_4$ controls global curvature and correlates closely with BFS. Average elevation maps can be used to validate the interpretation of clustering criteria identified through coefficient visualizations. When a clustering criterion aligns well with a clinical parameter, the cluster maps (the maps produced during the clustering process) and the clinical-parameter average elevation maps (or atlases) for the corresponding parameter show similar patterns. Clinical maps are constructed by averaging corneal elevations within groups defined by specific clinical parameters [4]. Here, the cluster maps show the same progression as the clinical-parameter average elevation maps for BFS [21,33]. However, it is worth noting that cluster maps and clinical maps may not be identical, as clinical maps are based on predefined subdivisions, whereas cluster maps reflect the natural distribution of the data. Statistical cluster comparisons linked the clusters to BFS [14,21], as well as to WTW and sex [14,22,23]. These findings highlight BFS as the main determinant of variation in corneal shape, with WTW and sex having smaller effects (Table 3). Notably, BFS radius was also identified as the main predictor of corneal shape in [10]. Moreover, in another study on a biometric system for person authentication based on corneal shape [34], the authors also identified $C_4$ as a much more discriminating feature than any other feature, using a completely different dataset and a different corneal topographer (Pentacam). These findings demonstrate that the clusters are related to significant clinical parameters, confirming that the algorithm correctly extracted relevant information even though it was based solely on the normal 3D shape of the corneas. It is interesting to note that in another study [33], clustering tests were conducted on BFS-normalized corneas to eliminate the effects of BFS size and global curvature variation, leaving only local features. In that case, $C_6$ (vertical astigmatism $Z_2^2$) emerged as the main clustering criterion and was related to another important clinical parameter: eye axis.

Regarding computational performance, the overall runtime of the complete k-means pipeline for computing the clusters of normal adult anterior corneal surfaces—including testing $k = 2$ to $k = 10$, computing scores, generating statistics, and producing figures—was less than 5 minutes on a typical Windows 10 Pro workstation.

## 5. Conclusion

This study investigated a large number of normal adult corneal topographies in an exploratory attempt to identify their natural groupings based on shape. To facilitate grouping, Zernike polynomial modeling of the elevation maps was performed prior to k-means clustering. The resulting clusters were evaluated using multiple metrics, with the optimal number of clusters being three (though four was also acceptable), primarily reflecting variations in corneal curvature.

In addition to a better knowledge of the shape of the human cornea, clustering of corneas also provides a powerful tool for establishing the optimal number of 3D models for the production of biosynthetic corneal substitutes. This methodology indicates which of three different categories would confer to the implant the shape that best corresponds to the anatomy of the patient's eye in order to rehabilitate corneal shape and function. It will allow a better adjustment to the recipient than the current one-size-fits-all solid implants, and it should also improve predictability compared to traditional eye bank donor corneas, for which no information on shape is available. The same methodology could be used for the categorization of the posterior surface and pachymetry maps individually or in combination with anterior surface for additional refinement of the implant. This innovative concept of a tailor-made corneal implant and its future optimization on different types of corneal pathologies promise a significant advance in the field of corneal transplantation.

## Acknowledgements

This work was supported by the Quebec Vision Health Research Network (VHRN), the FRQNT (Fonds de recherche du Québec Nature et Technologie) and the MUTAN (University Mission of Tunisia in North America).